\documentclass[journal]{IEEEtran}
\usepackage{cite}

\ifCLASSINFOpdf
\usepackage[pdftex]{graphicx}
\DeclareGraphicsExtensions{.pdf,.jpeg,.png}
\else
\usepackage[dvips]{graphicx}
\DeclareGraphicsExtensions{.eps}
\fi
\usepackage{epsfig,epsf,epstopdf,graphicx}
\usepackage[cmex10]{amsmath}
\usepackage{amssymb}
\usepackage{amsthm }
\usepackage{nicefrac}
\usepackage{hyperref}
\usepackage{xcolor}
\hypersetup{colorlinks=true}
\usepackage{tabularx}

\theoremstyle{theorem}

\theoremstyle{definition}

\theoremstyle{plain}

\theoremstyle{plain}

\newcommand\Tstrut{\rule{0pt}{2.6ex}}         
\newcommand{\kb}{{\textbf{k}}}

\newcommand{\db}{{\textbf{d}}}

\newcommand{\ab}{{\textbf{a}}}

\newcommand{\fb}{{\textbf{f}}}

\newcommand{\mb}{{\textbf{m}}}

\newcommand{\Ab}{{\textbf{A}}}
\newcommand{\nb}{{\textbf{n}}}

\newcommand{\Rb}{{\textbf{R}}}

\newcommand{\vb}{{\textbf{v}}}

\newcommand{\Pb}{{\textbf{P}}}

\newcommand{\pb}{{\textbf{p}}}

\newcommand{\etab}{{\mbox{\boldmath $\eta$}}}

\newcommand{\thetab}{{\mbox{\boldmath $\theta$}}}

\newcommand{\Wb}{\mathbf{W}}

\usepackage{algorithm}
\usepackage{multirow}
\usepackage{algcompatible}
\usepackage[bottom]{footmisc}

\PassOptionsToPackage{dvipsnames}{xcolor}
\usepackage{tikz}
\usetikzlibrary{calc}

\ifCLASSOPTIONcompsoc
  \usepackage[caption=false,font=normalsize,labelfont=sf,textfont=sf]{subfig}
\else
  \usepackage[caption=false,font=footnotesize]{subfig}
\fi
\usepackage{fixltx2e}
\usepackage{afterpage}
\usepackage{float}
\usepackage{stfloats}

\makeatletter

\begin{document}

%
\title{Closed-form Solutions for Velocity and Acceleration of a Moving Vehicle Using Range, Range Rate, and Derivative of Range Rate}


\author{Mohammad Salman, Hadi Zayyani, Hasan Abu Hilal, Mostafa Rashdan


\thanks{M.~Salman and M.~Rashdan are with College of Engineering and Technology, American University of the Middle East, Egaila, 54200, Kuwait (e-mails: mohammad.salman@aum.edu.kw, mostafa.rashdan@aum.edu.kw).}
\thanks{H.~Zayyani is with the Department
of Electrical and Computer Engineering, Qom University of Technology (QUT), Qom, Iran (e-mail: zayyani2009@gmail.com).}
\thanks{H.~Abu Hilal is with Electrical Engineering Department, Higher Colleges of Technology, Abu Dhabi, UAE (e-mail: hasan.abuhilal@hct.ac.ae).}

\vspace{-0.5cm}}


\maketitle
\thispagestyle{plain}
\pagestyle{plain}

\begin{abstract}
This letter presents a novel method for estimating the position, velocity, and acceleration of a moving target using range-based measurements. Although most existing studies focus on position and velocity estimation, the framework of this letter is extended to include acceleration. To achieve this, we propose using the derivative of the range rate, in addition to the range and range rate measurements. The proposed method estimates the position at first using Time-of-Arrival (TOA)-based techniques; then, develops a reformulated least squares (LS) and weighted least squares (WLS) approaches for velocity estimation; and finally, employs the derivative of the range rate to estimate the acceleration using previous position and velocity estimates. On the other hand, closed-form LS and WLS solutions are derived for both velocity and acceleration. The simulation results show that the proposed approach provides improved performance in estimating moving target kinematics compared to existing methods.
\end{abstract}

\begin{IEEEkeywords}
Localization, range, range rate, derivative, weighted least squares.
\end{IEEEkeywords}

%
\IEEEpeerreviewmaketitle

\section{Introduction}
\label{sec:Intro}
\IEEEPARstart{T}{he} problem of estimating the location and velocity of a moving target is an important topic in the fields of communication and signal processing. This critical issue arises in many applications, such as radar and sensor networks. This issue has been addressed using two broad categories of radar systems; active and passive radars, direct and indirect localization techniques. In this context, the focus will be on the general concept of indirect localization, which involves a two-stage process.

\noindent In the first stage, one or more features of the moving target are extracted from the received signal. These features include Received Signal Strength (RSS) \cite{Mahd21},  fingerprinting \cite{Homma25}, \cite{Cheng24}, Time of Arrival (TOA) \cite{Shi20}, \cite{Chen20} Time Difference of Arrival (TDOA) \cite{Moti24}, Frequency Difference of Arrival (FDOA) \cite{Ho04, Yu12, Noroo17}, Angle of Arrival (AOA) \cite{Arab23}, \cite{Yuan19}, and Doppler Shift (DS) \cite{Nguy18}, or using multiple of features \cite{Zia23}, and others. In the second stage, these measured features are processed to estimate the velocity and location of the target.

Motivated by the pursuit of algebraic closed-form solutions, using TDOA and FDOA measurements, some papers proposed closed-form solutions the position and velocity estimation of a target \cite{Ho04}, \cite{Yu12}, \cite{Noroo17}. In addition, an asymptotically efficient estimator was presented to determine the position and velocity of a moving target from Time Delay (TD) and DS measurements in the presence of location uncertainties \cite{Noroo19}. Moreover, in \cite{Li19}, a closed-form solution was presented for localization using TOA and Frequency of Arrival (FoA) measurements with a spherical constraint. Subsequently, the authors in \cite{Yang20} presented a three-dimensional (3D) joint estimation which utilizes bistatic range (BR), TDOA, and DS to improve the accuracy of position and velocity estimation of a moving target in a multistatic radar system. Moreover, Kazemi \textit{et al.} \cite{Kazemi20} proposed an Approximate Maximum Likelihood (ALM) approach for moving target localization in distributed MIMO radar systems using TD and DS measurements. Besides \cite{Liu21} proposed an efficient estimator for source localization Using TD and AOA Measurements in MIMO Radar Systems. In addition, Jabbari \textit{et al.} \cite{Jabbari23}, proposed a joint estimation method for the position and velocity of a moving target in distributed networks of moving radars using TOA and DS measurements. In the context of asynchronous MIMO systems, the authors in \cite{Wu24} proposed an optimization-based Semi-Definite Programming (SDP) method to obtain the position and velocity of a moving target by using Differential Time Delay (DTD) and Differential Frequency Shift (DFS).

\noindent Due to recent studies that extend the estimation beyond position and velocity and include the acceleration of a moving target \cite{Sun23, Jiang25}, this study also aims to estimate the acceleration of a moving target. To achieve this, we propose using the derivative of the range rate in addition to the range and the range rate. The proposed approach first estimates the location using a conventional TOA-based method; then, using a novel reformulation, LS and WLS solutions are derived for velocity estimation; finally, the acceleration is estimated independently by reformulating the derivative of the range rate, leveraging the previously estimated location and velocity. LS and WLS methods are also developed for this final step.

\noindent The main contributions of this letter are:
\begin{itemize}
\item Introducing the use of the derivative of range rate as an additional measurement for a moving target.
\item Reformulating the estimation problem and presenting LS and WLS solutions for velocity and acceleration.
\item Extension of traditional kinematic estimation by incorporating acceleration estimation in addition to position and velocity.
\end{itemize}
\noindent Simulation results show the superiority of the proposed WLS method over a state-of-the-art method in the literature in high Signal to Noise Ratios (SNR). Moreover, the complexity of the proposed algorithm in terms of simulation run time is less than others.


\section{System Model and Problem Formulation}
\label{sec:ProblemForm}
\noindent Consider a sensor network consisting of $N$ fixed sensors in a two-dimensional space. Each sensor can measure the range, the range rate and the derivative of the range rate of a moving vehicle. Let the position of the $i$'th sensor located in a 2D space be denoted $\pb_i=[x_i,y_i]^T$, and the position of the moving target at time $t$ be $\pb_{s,t}=[x_{s,t},y_{s,t}]^T$, with a velocity given by $\vb_{s,t}=[v_{sx,t},v_{sy,t}]^T$. The range between the $i$'th sensor and the vehicle is $r_i=||\pb_s-\pb_i||_2$ and can be measured, for example, by estimating the TOA. The set of ranges $r_i, 1\le i\le N$ is sufficient to estimate the vehicle's position using, for example, a trilateration method. However, estimating the velocity of the vehicle requires more information. To this end, the range rates defined as $\dot{r}_i=\frac{d}{dt}||\pb_s-\pb_i||$ are utilized.
These range rates provide information that can be leveraged to estimate the velocity of the vehicle, as will be discussed in the next section.

\noindent To further explore the properties of the moving vehicle, we propose utilizing the derivative of range rates (which is the second derivative of ranges) defined as $\ddot{r}_i=\frac{d}{dt^2}||\pb_s-\pb_i||$. Thus, considering the measurement noises, the system incorporates three groups of measurements, namely:
\begin{align}
\label{eq: meas}
&\bar{r}_i=r_i+v_i=||\pb_s-\pb_i||+v_i,\nonumber \\
&a_i=\dot{r}_i+n_i=\frac{d}{dt}||\pb_s-\pb_i||+n_i,\nonumber\\
&b_i=\ddot{r}_i+\eta_i=\frac{d}{dt^2}||\pb_s-\pb_i||+\eta_i,
\end{align}

\noindent where $v_i$, $n_i$, and $\eta_i$ are the measurement noises of the range, the range rate, and the derivative of the range rate, respectively. The objective is to estimate the position $\pb_s$, the velocity $\vb_s=\frac{d}{dt}\pb_s$ and the acceleration $\ab_s=\frac{d}{dt}\vb_s$  of the vehicle using these three noisy measurements; $\bar{r}_i$, $d_i$, and $b_i$ for $1\le i\le N$.

\section{The Proposed Algorithm for Velocity Estimation}
\label{sec: prop1}
\noindent To develop an algorithm to estimate the kinematic parameters of the moving target; namely, its location, velocity, and acceleration, from the measurements in (\ref{eq: meas}), we start by driving simplified expressions for these measurements. The starting point is to derive the expression for the range rate, which can be obtained as
\begin{align}
\label{eq: rr}
a_i=\dot{r}_i+n_i=\frac{\vb^T_s(\pb_s-\pb_i)}{||\pb_s-\pb_i||}+n_i.
\end{align}

\noindent Taking the time derivative of (\ref{eq: rr}) and with some algebraic manipulations, yields the following result
\begin{align}
\label{eq: drr}
b_i&=\ddot{r}_i+\eta_i=\frac{d}{dt}\Big[\frac{\vb^T_s(\pb_s-\pb_i)}{||\pb_s-\pb_i||}\Big]\nonumber\\
&=\frac{\frac{d}{dt}\Big[\vb^T_s(\pb_s-\pb_i)\Big]||\pb_s-\pb_i||-\vb^T_s(\pb_s-\pb_i)\frac{d}{dt}||\pb_s-\pb_i||}{||\pb_s-\pb_i||^2}\nonumber\\
&\approx\frac{\Big(\frac{d\vb_s}{dt}\Big)(\pb_s-\pb_i)+||\vb_s||^2-a_i\vb^T_s(\pb_s-\pb_i)}{||\pb_s-\pb_i||^2}+\eta_i,
\end{align}

\noindent where the approximation in (\ref{eq: drr}) is due to the neglect of the noise term $n_i$ in (\ref{eq: rr}) when substituting $a_i$ in place of $\frac{d}{dt}||\pb_s-\pb_i||=\frac{\vb^T_s(\pb_s-\pb_i)}{||\pb_s-\pb_i||}$.

\noindent Since the three equations for the range, the range rate and the derivative of the range rate are interconnected and quite complex to jointly estimate the position, velocity, and acceleration, we adopt a sequential approach to estimate these parameters independently. We begin with TOA-based equations to estimate the position $\pb_s=[x_s,y_s]^T$ using only the range measurements. This is a classical approach in TOA-based localization. This results in
\begin{align}
\hat{\thetab}=\Ab^{\dagger}\fb,
\end{align}

\noindent where $\thetab=[x_s,y_s,x^2_s+y^2_s]^T$, $^{\dagger}$ is the pseudoinverse operator, $f_i=r^2_i-x^2_i-y^2_i$ and $\Ab(i,:)=[-2x_i,-2y_i,1]$ for $1\le i\le N$.

\subsection{LS estimation of velocity}
\label{sec: LS}
\noindent After estimating the position of the moving target, we proceed to estimate the velocity of the moving target using the range rates $a_i$'s for all $1\le i\le N$. Thus, reformulating (\ref{eq: rr}) by multiplying its two sides by $||\pb_s-\pb_i||$, yields
\begin{align}
\label{eq: nc}
\db=\mathbf{1}_N\vb^T_s\pb_s-\Pb^T\vb_s+\nb_c,
\end{align}

\noindent where $\db=[d_1,d_2,...,d_N]^T$ is an $N\times 1$ vector, $d_i=a_ir_i$ is a scalar, $\Pb=[\pb_1|\pb_2|...|\pb_N$ is a $2\times N$ matrix, and $\nb_c=[n_{c,1},...,n_{c,N}]^T$, $n_{c,i}=r_in_i$. LS method is used to estimate the velocity vector $\vb_s$ using the vector $\db$. The LS cost function is defined as
\begin{align}
\label{eq: J}
\mathrm{J}(\vb_s)=||\db-\mathbf{1}_N\vb^T_s\pb_s+\Pb^T\vb_s||^2_2=||\db-\ab||^2,
\end{align}

\noindent where $\ab=\mathbf{1}_N\vb^T_s\pb_s-\Pb^T\vb_s$. Neglecting the constant terms of $\mathrm{J}(\vb_s)$ in (\ref{eq: J}) provides

\begin{align}
\label{eq: JJ}
\mathrm{J}(\vb_s)&\equiv -2\db^T\ab+\ab^T\ab.
\end{align}

\noindent The term $\ab^T\ab$ in (\ref{eq: JJ}) can be simplified as
\begin{align}
\label{eq: Q}
\ab^T\ab &=(\mathbf{1}_N\vb^T_s\pb_s-\Pb^T\vb_s)^T(\mathbf{1}_N\vb^T_s\pb_s-\Pb^T\vb_s)\nonumber\\
&=N\vb^T_s\pb_s\pb^T_s\vb_s-2\vb^T_s\Pb\mathbf{1}_N\pb^T_s\vb_s+\vb^T_s\Pb\Pb^T\vb_s\nonumber\\
&=\vb^T_s\Rb^{*}\vb_s,
\end{align}

\noindent where $\Rb^{*}=N\Rb_{ps}-2\Rb_0+\Rb_p$, where $\Rb_{ps}\triangleq \pb_s\pb^T_s$, $\Rb_0\triangleq \Pb\mathbf{1}_N\pb^T_s$, and $\Rb_p\triangleq \Pb\Pb^T$. Now, substituting (\ref{eq: Q}) into (\ref{eq: JJ}) yields
\begin{align}
\label{eq:cost}
\mathrm{J}(\vb_s)\equiv-2\db^T\mathbf{1}_N\vb^T_s\pb_s+2\db^T\Pb^T\vb_s+\vb^T_s\Rb^{*}\vb_s.
\end{align}

\noindent The cost function in (\ref{eq:cost}) is quadratic in $\vb_s$. Taking the derivative with respect to $\vb_s$ and setting it to zero; $\frac{\partial \mathrm{J}(\vb_s)}{\partial \vb_s}=\mathbf{0}$, provides the following closed-form solution for velocity estimation:
\begin{align}
\hat{\vb}_{s,LS}=c\Big(\Rb^{*}+\Pb\db\Big)^{-1}\pb_s,
\end{align}

\noindent where $c\triangleq \db^T\mathbf{1}_N$.

\subsection{WLS estimation of velocity}
\noindent Since the noise term in (\ref{eq: nc}) is range dependent, a WLS method is more convenient to estimate the velocity. The weights can be defined as $W_i=\frac{1}{||\hat{\pb}_s-\pb_i||}$, in which the measurements with the higher associated noise are assigned lower weights. This WLS approach is an iterative method in nature to assign initial weights and then estimate the velocity, and then the process is repeated.

\noindent The WLS cost function is then defined as
\begin{align}
\label{eq: WLS}
\mathrm{J}_{\mathrm{WLS}}(\vb_s)&=(\db-\ab)^T\Wb(\db-\ab)\nonumber\\
&\equiv -2\db^T\Wb\ab+\ab^T\Wb\ab,
\end{align}
\noindent where $\Wb=\mathrm{diag}(W_i)$is the diagonal weighting matrix. Following a similar derivation as presented in Subsection \ref{sec: LS}, we obtain $\ab^T\Wb\ab=\vb^T_s\Rb^{*}_w\vb_s$, where $\Rb^{*}_w=\Rb_{ps,w}-2\Rb_{0,w}+\Rb_{pw}$ in which $\Rb_{ps,w}\triangleq \pb_s\mathbf{1}^T_N\Wb\mathbf{1}_N\pb^T_s$, $\Rb_{0,w}\triangleq \Pb\Wb\mathbf{1}_N\pb^T_s$, and $\Rb_{pw}\triangleq \Pb\Wb\Pb^T$. Again, some calculations result to the WLS cost function as $\mathrm{J}_{\mathrm{WLS}}(\vb_s)\equiv-2\db^T\Wb(\mathbf{1}_N\vb^T_s\pb_s-\Pb^T\vb_s)+\vb^T_s\Rb^{*}_w\vb_s$. Similarly, the minimum of the WLS cost function is obtained by setting its partial derivative with respect to $\vb_s$ equal to zero. Thus, the WLS-based velocity estimator is given as
\begin{align}
\hat{\vb}_{s,WLS}=c_w\Big(\Rb^{*}_w+\Pb\Wb^T\db\Big)^{-1}\pb_s,
\end{align}

\noindent where $c_w\triangleq \db^T\Wb\mathbf{1}_N$.

\section{The Proposed Algorithm for Acceleration Estimation}
\label{sec: prop2}
\noindent In this section, the acceleration of a moving target is estimated using the derivative of the range rates $b_i$'s, using the previously estimated location and velocity obtained in Section~\ref{sec: prop1}. By defining the acceleration vector as $\ab_s\triangleq \frac{d\vb_s}{dt}$, and using the derivative of the range rates in (\ref{eq: drr}), the following approximation is obtained
\begin{align}
b_i\approx\frac{\ab^T_s(\pb_s-\pb_i)}{||\pb_s-\pb_i||^2}+\frac{||\vb_s||^2}{||\pb_s-\pb_i||^2}-\frac{a_i}{||\pb_s-\pb_i||}+\eta_i.
\end{align}

\noindent Given that $\pb_s$ and $\vb_s$ are already estimated, this expression can be written as
\begin{align}
\label{eq: btild}
\tilde{b}_i=b_i-\frac{||\vb_s||^2}{||\pb_s+\pb_i||^2}-\frac{a_i}{||\pb_s-\pb_i||}\approx\frac{\ab^T_s(\pb_s-\pb_i)}{||\pb_s-\pb_i||^2}+\eta_i.
\end{align}

\noindent Multiplying both sides of (\ref{eq: btild}) by $||\pb_s-\pb_i||^2$, yields
\begin{align}
\label{eq: ki}
k_i\triangleq \tilde{b}_i||\pb_s-\pb_i||^2=\ab^T_s(\pb_s-\pb_i)+\eta^{'}_i,
\end{align}

\noindent where $\eta^{'}_i=\eta_i||\pb_s-\pb_i||^2$.

\subsection{LS estimation of acceleration}
\noindent To estimate the acceleration vector $\ab_s$ using the transformed derivative of the range rate in (\ref{eq: ki}), we collect all the measurements of $k_i$ in a vector $\kb\triangleq [k_1,k_2,...k_N]^T$. Following similar steps as in the previous sections, the transformed measurement vector $\kb$ can be written as
\begin{align}
\kb=\mathbf{1}_N\ab^T_s\pb_s-\Pb^T\ab_s+\etab^{'},
\end{align}

\noindent where $\etab^{'}=[\eta^{'}_1,...,\eta^{'}_N]^T$. Such a system can be solved using LS and hence the following LS cost function can be used:
\begin{align}
\mathrm{G}(\ab_s)=||\kb-\mathbf{1}_N\ab^T_s\pb_s+\Pb^T\ab_s||^2_2=||\db-\ab^{'}||^2,
\end{align}

\noindent where $\ab^{'}\triangleq\mathbf{1}_N\ab^T_s\pb_s-\Pb^T\ab_s$. Following similar derivations, which are omitted for brevity, the final LS estimator for the acceleration is written as
\begin{align}
\hat{\ab}_{s,LS}=c_a\Big(\Rb^{*}_a+\Pb\kb\Big)^{-1}\pb_s,
\end{align}
\noindent where $c_a\triangleq \kb^T\mathbf{1}_N$, $\Rb^{*}_a=\Rb^{*}$.

\subsection{WLS estimation of acceleration}
\noindent For WLS estimation, the WLS cost function is similarly equal to
\begin{align}
\mathrm{J}_{\mathrm{WLS}}(\ab_s)\equiv-2\kb^T\Wb(\mathbf{1}_N\ab^T_s\pb_s-\Pb^T\ab_s)+\ab^T_s\Rb^{*}_{w,a}\ab_s,
\end{align}

\noindent where $\Rb^{*}_{w,a}=\Rb^{*}_w$. Similar steps as before, the final WLS estimator for the acceleration is equal to
\begin{align}
\hat{\ab}_{s,WLS}=c_{w,a}\Big(\Rb^{*}_w+\Pb\Wb^T\kb\Big)^{-1}\pb_s,
\end{align}

\noindent where $c_{w,a}\triangleq \kb^T\Wb\mathbf{1}_N$.

\section{Simulation Results}
\label{sec:SimulationDiscussion}
In this section, the performance of the proposed LS and WLS algorithm is investigated. The number of sensors is considered as $N=8$ and the sensors are placed in a square 2D region of $[-100, 100]\times[-100,100]$. The sensor positions are $[0,0]^T$, $[100,100]^T$, $[-100,100]^T$, $[100,-100]^T$, $[-100,-100]^T$, $[-50,50]^T$, $[50,50]^T$, and $[-50,-50]^T$. The initial position of moving vehicle is uniformly distributed $[0,100]\times[0,100]$. Two experiments were performed. The first experiment is the case of the constant velocity of the vehicle. The constant velocity elements of $\vb_s$ are chosen uniformly random in $[-20,20]$. The second experiment is the case of the constant acceleration of the vehicle. In this case, the initial position is selected as in the first experiment. The constant acceleration is selected uniformly in $[-10,10]$. The errors of range, range gate, and derivative of range rate is assumed to be Gaussian with zero mean and default variances equal to 1, otherwise stated. The performance metric is defined as Root-Mean-Square-Error of position estimator, velocity estimator, and acceleration estimator defined as $\mathrm{RMSE}(\hat{\mb})=\sqrt{\frac{1}{K}\sum_{k=1}^K||\mb-\hat{\mb}||_2^2}$, where $\mb$ is the position, velocity, or acceleration, $\hat{\mb}$ is the estimated position, velocity, or acceleration, and the averaging is done over $K=1000$ different random Monte Carlo simulations. In the first experiment, three algorithms and a Cramer-Rao-Lower-Bound (CRLB) are simulated and the results are compared. The two proposed methods are proposed LS and proposed WLS. The competing method is Quasi-Synchronous Semi-Definite Programming (QS-SDP) presented in \cite{Wu24} in which differential time delay (DTD) and differential frequency shift (DFS) measurements are used to joint position and velocity estimation. Also, for a benchmark comparison, we used the CRLB computed in \cite{Wu24}.

\begin{figure}[ht]
\begin{center}
\includegraphics[width=5.5cm]{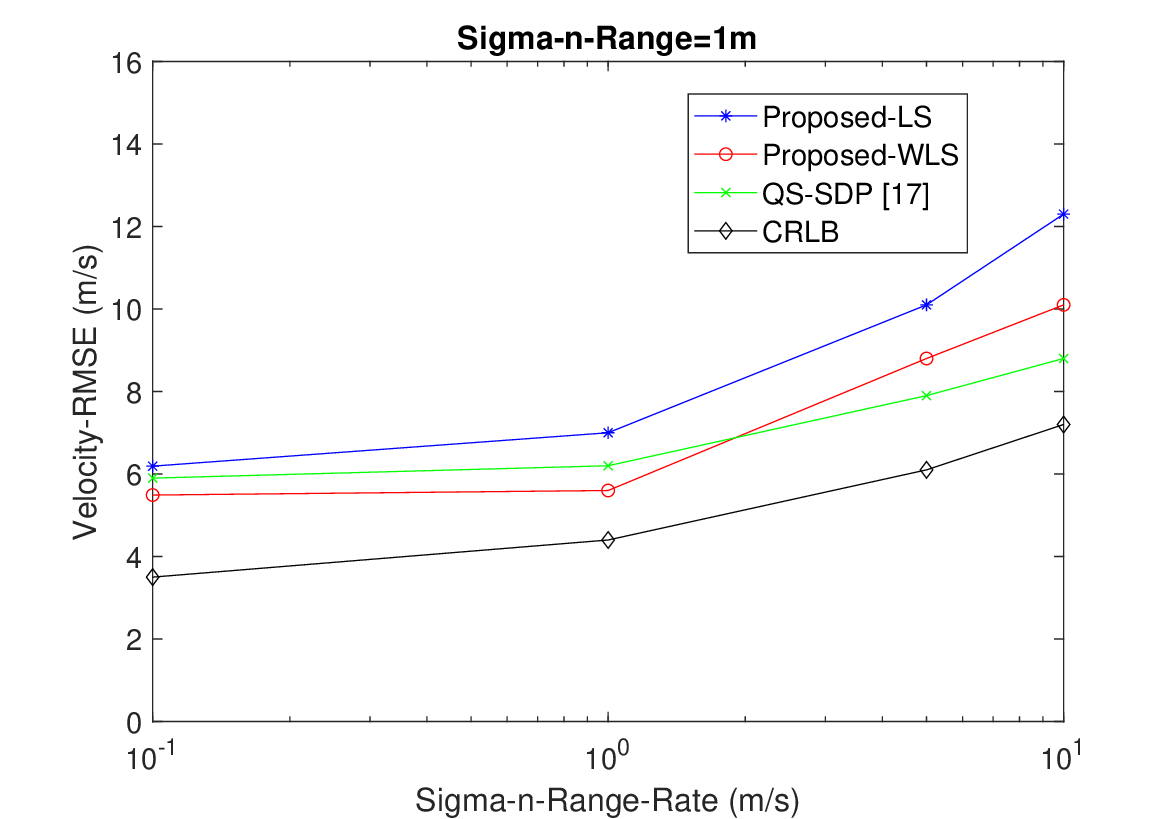}
\end{center}
\vspace{-5mm}
\caption{Velocity RMSE versus standard deviation of range rate error.}
\label{fig3}
\end{figure}

In the first experiment of constant velocity vehicle, a noise analysis of range rate error was performed. In this case, we choose a fixed standard deviation of range error equal to 1m, and the standard deviation of range rate error is varied between 0.1 to 10. The results of velocity RMSE versus the standard deviation of range rate error is shown in Figure.~\ref{fig3}. It shows that the proposed WLS method outperforms QS-SDP when standard deviation is less than 2m/s. Also, the position RMSE is almost constant and is not depicted for brevity.


\begin{table}[!b]
\caption{Complexity in terms of average simulation run time of various algorithms}
\centering
 \begin{tabular}{p{32mm}||p{11mm}|p{11mm}|p{11mm}} \hline
 \Tstrut & Proposed LS& Proposed WLS& QS-SDP \\\hline \hline

 \Tstrut Average run time (seconds)
 & $\!\!\! \begin{aligned} 0.15& \end{aligned} $	
 & $\!\!\!\begin{aligned} 0.18&\end{aligned}$	
 &	$\!\!\!\begin{aligned} 1.12&\end{aligned}$  \\ \hline

%
%
%
\end{tabular}
 \begin{tabular} {l}
\end{tabular}
\label{Table_1}
\end{table}

\begin{figure}[ht]
\begin{center}
\includegraphics[width=5.5cm]{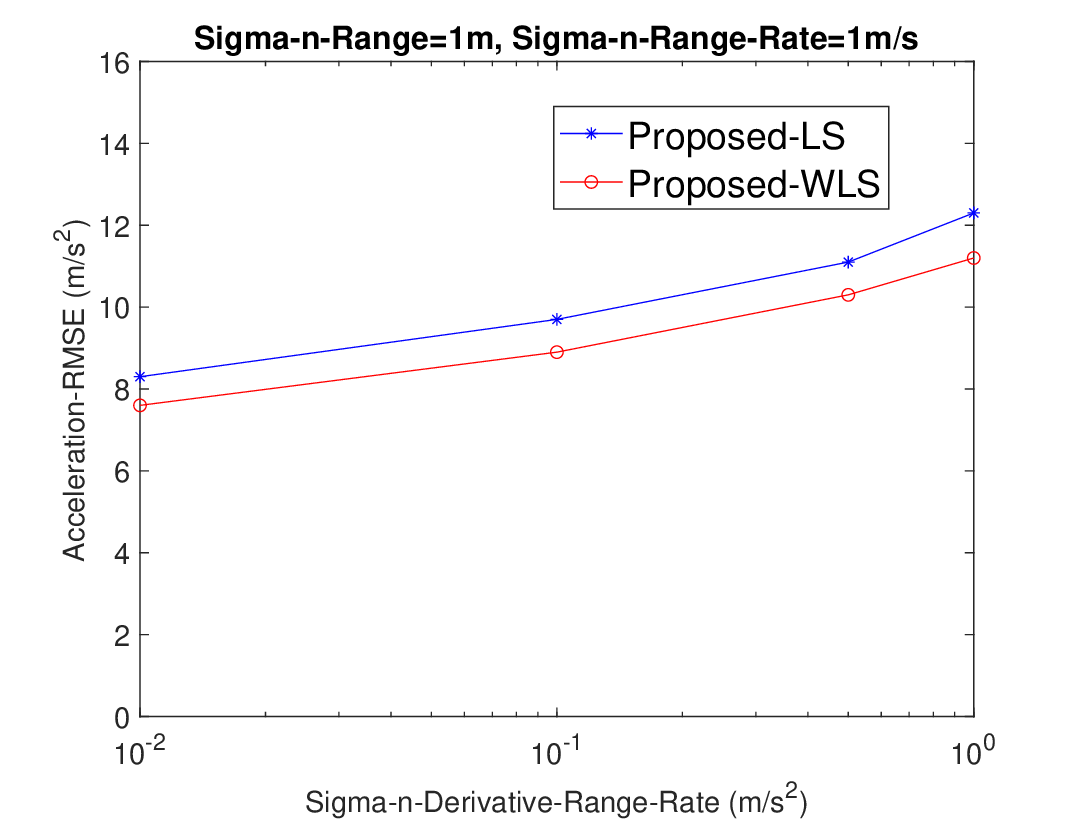}
\end{center}
\vspace{-5mm}
\caption{Acceleration RMSE versus standard deviation of derivative of range rate error.}
\label{fig4}
\end{figure}

In the second experiment, acceleration estimation is investigated. In this case, the standard deviation of range error and range rate error are selected as 1 and the standard deviation of derivative of range rate is varied between 0.01 to 1. The results of acceleration RMSE versus the standard deviation of derivative of range rate is demonstrated in Figure.~\ref{fig4}. It shows that the proposed WLS is better than the proposed LS.

To compare the complexity of algorithms, we use the average simulation run time of algorithms which are depicted in Table~1. It shows that the least complex algorithm is the proposed LS and WLS algorithm in comparison to QS-SDP.

\section{Conclusion}
\label{sec:conc}
This letter presents a method for velocity and acceleration estimation of a moving vehicle using range measurements, range rate measurements, and derivative of range rates. Two LS and WLS approaches are suggested to solve the problem by reformulating the measurement equations of range rate and of range rate measurements. The simulation results show the advantage of the proposed methods especially WLS method in the low noise conditions. Moreover, the computational complexity of the proposed methods in terms of simulation run time is one order of magnitude less than the competing method.

\end{document}